\documentclass[english,prl, twocolumn, showpacs]{revtex4}
\usepackage[T1]{fontenc}
\usepackage[latin9]{inputenc}
\usepackage{array}
\usepackage{multirow}
\usepackage{amstext}
\usepackage{amssymb}
\usepackage{graphicx}
\usepackage{esint}

\makeatletter

\providecommand{\tabularnewline}{\\}

\@ifundefined{textcolor}{}
{%
 \definecolor{BLACK}{gray}{0}
 \definecolor{WHITE}{gray}{1}
 \definecolor{RED}{rgb}{1,0,0}
 \definecolor{GREEN}{rgb}{0,1,0}
 \definecolor{BLUE}{rgb}{0,0,1}
 \definecolor{CYAN}{cmyk}{1,0,0,0}
 \definecolor{MAGENTA}{cmyk}{0,1,0,0}
 \definecolor{YELLOW}{cmyk}{0,0,1,0}
 }

\makeatother

\usepackage{babel}
\begin{document}

\title{Superradiant quantum phase transition in a circuit QED system: \\a
revisit from a fully microscopic point of view}

\author{D. Z. Xu$^{1}$, Y. B. Gao$^{2}$, and C. P. Sun$^{1}$}

\email{suncp@itp.ac.cn}

\homepage{http://power.itp.ac.cn/ suncp/index.html}

\affiliation{$^{1}$State Key Laboratory of Theoretical Physics,Institute of Theoretical
Physics, Chinese Academy of Science, Beijing 100190, People's Republic
of China\\
 $^{2}$College of Applied Science, Beijing University of Technology,
Beijing 100124, China}
\begin{abstract}
In order to examine whether or not the quantum phase transition of
Dicke type exists in realistic systems, we revisit the model setup
of the superconducting circuit QED from a microscopic many-body perspective
based on the BCS theory with pseudo-spin presentation. By deriving
the Dicke model with the correct charging terms from the minimum coupling
principle, it is shown that the circuit QED system can exhibit superradiant
quantum phase transition in the limit $N\rightarrow\infty$. The critical
point could be reached at easiness by adjusting the extra parameters,
the ratio of Josephson capacitance $C_{J}$ to gate capacitance $C_{g}$,
as well as the conventional one, the ratio of Josephson energy $E_{J}$
to charging energy $E_{C}$.
\end{abstract}


\maketitle
\textit{Introduction.}\textemdash{}In the conventional Dicke model
\cite{Dicke1954}, which ignores two-photon interaction term $\mathbf{A}^{2}$,
the superradiant quantum phase transition (QPT) \cite{Lieb1973,EmaryPRL,YLi2006}
can happen when the atom-radiation field coupling $g$ is strong enough.
However, for the realistic systems with the minimum coupling from
$U\left(1\right)$-gauge theory, the Thomas-Reiche-Kuhn (TRK) sum
rule means that the $\mathbf{A}^{2}$ term cannot be neglected when
the increases of $\mathbf{A}^{2}$ term follows the increase of $g$
\cite{Rza=00017Cewski1975,Zaffino2004}. The $\mathbf{A}^{2}$ term
shifting the effective frequencies hence prevents the considered system
from reaching the critical point, so that no superradiant QPT happens
in the natural atom systems. This fact was stated as a no-go theorem
for the realistic cavity QED systems.

On the contrary, it was ad hoc pointed out that this superradiant
QPT could be realized in some artificial system\cite{Sun2009}. Later
on, P. Nataf and C. Ciuti showed that the circuit QED system consisting
of a collection of Josephson atoms capacitively coupled to a transmission
line resonator (TLR) is capable for such kind of QPT \cite{Nataf2010}.
Viehmann \textit{et al} questioned this judgment based on an overall
microscopic model. They argued \cite{Marquardt2011,Nataf2011,Marquardt2012}
that the phenomenological Hamiltonian used in Ref.\cite{Nataf2010}
cannot adequately describe the superradiant QPT of the circuit QED
system with large atom numbers; if all the degrees of freedom are
considered properly, the no-go theorem of superradiant QPT based on
the TKR sum rule still works for this artificial system. In this sense,
they excluded the existence of the superradiant QPT in the circuit
QED system. 

Viehmann \textit{et al} claimed that a fully microscopic approach
was utilized by themselves, but it seems difficult to straightforwardly
deduce the phenomenological Hamiltonian used in Ref.\cite{Nataf2010}
from their overall microscopic model. In this paper, we try to carry
out this necessary task to deduce it from a microscopic model with
the minimum coupling form $U\left(1\right)$-gauge theory. To this
end, we provide a description of the circuit QED system using the
pseudo-spin representation \cite{PALee1971} of the BCS theory, which
explicitly displays the superconducting characteristics of the Josephson
atoms. Our microscopic approach correctly gives the additional quadratic
quantum voltage term, which is usually ignored in current references,
e.g., \cite{Makhlin2001}. Applying this result to the low excited
ensemble of artificial atoms, we conclude that the no-go theorem in
the cavity QED system could not rule out the superradiant QPT in the
circuit QED system for some experimentally accessible parameters.

\begin{figure}
\includegraphics[width=7cm]{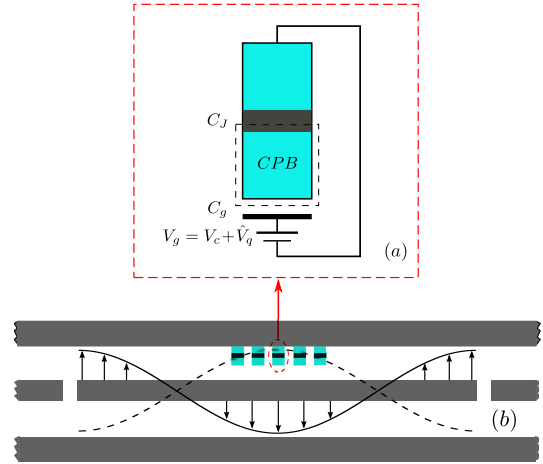}

\caption{\label{fig:geometry}(color online) The schematic of the superconducting
circuit for superradiant QPT. An array of Josephson junctions {[}with
zoomed picture in (a){]} is embedded in a superconducting TLR and
couples to the resonator mode capacitively (b). The equivalent circuit
for a single Josephson junction is depicted in (a), where the quantized
resonator mode $\hat{V}_{q}$ is included in the gate voltage $V_{g}$.}
\end{figure}

\textit{Microscopic modeling of superconducting circuit QED.}\textemdash{}As
a key element in the circuit QED system, as illustrated in Fig.$\ref{fig:geometry}$,
superconducting Josephson junction consists of a thin insulating barrier
sandwiched between two superconductors. Microscopically, we use the
collective pseudo-spin operators
\begin{eqnarray}
S_{\alpha} & = & \frac{1}{\sqrt{2}\mathcal{N}}\sum_{\mathbf{k}}c_{-\mathbf{k},\alpha}c_{\mathbf{k},\alpha},\nonumber \\
S_{z,\alpha} & = & \frac{1}{2}\sum_{\mathbf{k}}\left(c_{\mathbf{k},\alpha}^{\dagger}c_{\mathbf{k},\alpha}+c_{-\mathbf{k},\alpha}^{\dagger}c_{-\mathbf{k},\alpha}-1\right)\label{eq:Sdef-1}
\end{eqnarray}
to describe the charging (tunneling) process of Cooper pairs in the
junction \cite{Anderson1958,Stavn1965}. Here, $c_{\mathbf{k},\alpha}^{\dagger}\left(c_{\mathbf{k},\alpha}\right)$
is the electron creation (annihilation) operator for the superconductor
on the $\alpha$-hand side with $\alpha=L,R$. The index $\mathbf{k}\left(-\mathbf{k}\right)$
denotes the momentum of the electron with spin up (down), and it is
summed over the energy shell $\left[\epsilon_{F}-\hbar\omega_{D},\epsilon_{F}+\hbar\omega_{D}\right]$,
which is around the Fermi energy $\epsilon_{F}$ up to the Debye frequency
$\omega_{D}$. The normalization factor $\mathcal{N}$ equals to half
of the number of momentum states within this energy shell. Obviously,
$S_{z,\alpha}$ is the operator counting the number of the Cooper
pairs in excess of the electroneutrality of the superconductor on
the $\alpha$ side.

To model the tunneling process as the Josephson effect \cite{PALee1971},
the single electron tunneling Hamiltonian $H_{T}=T\sum_{\mathbf{k},\mathbf{q}}\left(c_{\mathbf{k},R}^{\dagger}c_{\mathbf{q},L}+h.c.\right)$
is re-expressed in terms of the collective pseudo-spin operators as
\begin{equation}
H_{T}=-2\hbar\mathcal{N}^{2}T\left(S^{+}+S^{-}\right),\label{eq:HT}
\end{equation}
where we have ignored the single electron tunneling terms for the
system in superconducting phase. Here, the operator $S^{-}=S_{L}S_{R}^{\dagger}$
($S^{+}=(S^{-})^{\dagger}$) denotes that a Cooper pair tunnels from
the left (right) superconductor to the right (left), where the operator
$S_{z}=\left(S_{z,L}-S_{z,R}\right)/2$ is defined and the commutation
relations $\left[S^{\pm},S_{z}\right]=\mp S^{\pm}$ and $\left[S^{+},S^{-}\right]\approx2S_{z}/\mathcal{N}^{2}$
are fulfilled. Note that $S_{z,L}+S_{z,R}=0$ since the Josephson
junction is electroneutral, $S_{z}$ also represents the number of
the excess Cooper pairs on the left bulk of superconductor. In the
case of small number of the excess tunneling Cooper pairs, i.e., $\mathcal{N}\gg\left\langle S_{z}\right\rangle $,
$S^{+}S^{-}$ is central to the algebra generated by $S^{\pm}$ and
$S_{z}$ because of $\left[S^{+},S^{-}\right]\rightarrow0$ \cite{Lugiato1980,Rodrigues2007}.
Then the polar decomposition $S^{\pm}=\exp\left(\pm i\theta\right)$
defines the macroscopic phase operator $\theta,$ which obeys $\left[\theta,S_{z}\right]=i.$
Using the phase operator, we can rewrite the tunneling Hamiltonian
in the conventional fashion $H_{T}=-E_{J}\cos\theta,$ where $E_{J}=\Phi_{0}I/2\pi$
is the Josephson energy, $\Phi_{0}=h/2e$ is the quantized flux, and
$I=8e\mathcal{N}^{2}T$ is the maximum tunneling current. 

According to the reference \cite{PALee1971}, the microscopic meaning
of $\theta=\phi_{L}-\phi_{R}$ could be explained as the difference
between the order parameters of the right- and left-hand superconductors
in the BCS ground states $\left\vert \phi_{\alpha}\right\rangle =\prod_{\mathbf{k}}\left[u_{\mathbf{k}}+v_{\mathbf{k}}\exp(i\phi_{\alpha})c_{-\mathbf{k},\alpha}^{\dagger}c_{\mathbf{k},\alpha}^{\dagger}\right]\left\vert 0\right\rangle $,
where $\phi_{\alpha}$ is the common phase of superconductor. Then
the tunneling current $\left\langle J\right\rangle =I\sin\left(\phi_{L}-\phi_{R}\right)$
is obtained by the average of $J=-2e\left[N_{L},H_{T}\right]$ over
the product state $\left\vert \phi\right\rangle =\left\vert \phi_{L}\right\rangle \otimes\left\vert \phi_{R}\right\rangle $,
where $N_{L}$ is the number of the electrons on the left-hand superconductor
which equals to $2S_{z}$ plus a constant.

Next we model the charging process for a simple Josephson device,
which is a superconducting island {[}or Cooper pair box (CPB){]} connected
to a gate capacitor $C_{g}$ and a bulk of superconducting electrode
through a thin junction with capacitance $C_{J\text{ }}$. The geometry
of the superconducting circuit is shown in Fig.$\ref{fig:geometry}$(a),
where the voltage $V_{c}$ is applied to the gate capacitor by a classical
source. This device is also coupled to a quantized electromagnetic
field provided by a superconducting TLR in a coplanar-waveguide geometry,
which gives an additional quantum voltage $\hat{V}_{q}=V_{q}\left(a+a^{\dagger}\right)$
where $a^{\dagger}\left(a\right)$ is the corresponding creation (annihilation)
operator for the single mode of the TLR with eigenfrequency $\omega_{r}$.

With the charges $Q$ distributed on the island, the electrostatic
potential $V$ of the junction is determined by 
\begin{equation}
C_{J}V-C_{g}\left(V_{g}-V\right)=Q\label{eq:potential}
\end{equation}
with the total gate voltage $V_{g}=V_{c}+\hat{V}_{q}.$ Initially,
we assume no excess electron exists, i.e., $Q=0$, hence the potential
of the CPB $V_{0}=C_{g}V_{g}/\left(C_{g}+C_{J}\right)$ is formally
quantized as an operator. The total electrostatic energy $U=C_{g}\left(V_{0}-V_{g}\right)^{2}/2+C_{J}V_{0}^{2}/2$
for both the gate and the Josephson capacitors connected to the electroneutral
CPB is calculated as 
\begin{equation}
U=4E_{C}n_{g}^{2}\frac{C_{J}}{C_{g}},\label{eq:U-1}
\end{equation}
where $E_{C}=e^{2}/2C_{\Sigma}$ is the charging energy for a single
electron, $C_{\Sigma}=C_{J}+C_{g}$ is the total capacitance and $n_{g}=C_{g}V_{g}/2e$.

After $l$ excess electrons are added in the CPB, the total energy
is the electrostatic energy $U$ plus the work $W$ done to tunneling
Cooper pairs. $W$ is actually the work cost by the excess electrons
to cross the barrier, which is actually supplied by the voltage source.
The corresponding potential $V_{l}$ is calculated by substituting
the excess charges $Q=Q_{l}\equiv-e\sum_{j=0}^{l}n_{j}$ into Eq.$\left(\ref{eq:potential}\right)$,
where $n_{j}=c_{j}^{\dagger}c_{j}$ is the single electron number
operator. As we concern the charge accumulation process, the momentum
states indexes $\mathbf{k}$ of the electron operators are of no importance.
Instead, we assign to each electron operator a subscript $j$ indicating
the order of accumulating on the island. 

According to classical electrodynamics, to add one more electron on
the island with $l$ excess electrons already on it, the work is calculated
according to the formula $W_{l}=-en_{l+1}V_{l}$. This formular seems
phenomenological, but now we can derive it from the minimum coupling
principle based on $U\left(1\right)$-gauge theory with a single particle
Hamiltonian $H_{e}=\left[\mathbf{p}-e\mathbf{A}\left(\mathbf{x}\right)\right]^{2}/2m-e\phi\left(\mathbf{x}\right)$.
It describes an electron moving in the vector potential $\mathbf{A}\left(\mathbf{x}\right)$
and scalar potential $\phi\left(\mathbf{x}\right)$. Here we use the
coulomb gauge $\mathbf{\mathbf{\nabla}}\cdot\mathbf{A}\left(\mathbf{x}\right)=0$
and the dipole approximation with $\mathbf{A}\left(\mathbf{x}\right)\approx\mathbf{A}\left(\mathbf{x}_{0}\right)\equiv\mathbf{A}_{0}$
and $\nabla\phi\left(\mathbf{x}\right)\approx\nabla\phi\left(\mathbf{x}\right)|_{x=x_{0}}$,
which is consistent with the prerequisite of the discussion about
the superradiant phenomenon in this paper. It leads to 
\begin{equation}
H_{e}\approx\frac{1}{2m}\left(\mathbf{p}^{2}+e^{2}\mathbf{A}_{0}^{2}\right)+e\mathbf{x}\cdot\mathbf{E}_{0},\label{eq:Hamil}
\end{equation}
where $\mathbf{E}_{0}=-\left[\partial\mathbf{A}\left(\mathbf{x}\right)/\partial t+\nabla\phi\left(\mathbf{x}\right)\right]|_{x=x_{o}}$. 

In second quantization, the field operators $\hat{\psi}\left(\mathbf{x}\right)=\sum_{\mathbf{k}}\psi_{\mathbf{k}}\left(\mathbf{x}\right)c_{\mathbf{k}}$
is used with $\psi_{\mathbf{k}}\left(\mathbf{x}\right)$ approximately
being the plane wave. Then the energy cost of a single electron crossing
the capacitor from one electrodes to the other one at $\mathbf{d}$
apart is calculated as 
\[
\int d^{3}x\left[\hat{\psi}^{\dagger}\left(\mathbf{x}\right)H_{e}\hat{\psi}\left(\mathbf{x}\right)\!-\!\hat{\psi}^{\dagger}\left(\mathbf{x}+\mathbf{d}\right)H_{e}\hat{\psi}\left(\mathbf{x}+\mathbf{d}\right)\right]=e\hat{n}V,
\]
where $V=\mathbf{d}\cdot\mathbf{E}_{0}$ and $\hat{n}=\sum_{\mathbf{k}}c_{\mathbf{k}}^{\dagger}c_{\mathbf{k}}$.
The momentum term $\mathbf{p}^{2}$ and quadratic vector potential
term $\mathbf{A}_{0}^{2}$ are both canceled out, and the remaining
term $-e\hat{n}V$ verifies our phenomenological formular of $W_{l}$.

When $N$ excess electrons are added in the CPB, the total work $W=\sum_{l=0}^{N-1}W_{l}$
is obtained as 
\begin{equation}
W=4E_{C}S_{z}^{2}-8E_{C}n_{g}S_{z}-2E_{C}S_{z},\label{eq:W-1}
\end{equation}
where we use the fact $S_{z}=\sum_{j=0}^{N}n_{j}/2$. The linear term
$2E_{C}S_{z}$ can be neglected because it will merely shift $n_{g}$
by $1/4$, which can be adjusted by tuning the gate voltage without
influence the further discussion. 

At last, the charging Hamiltonian $H_{C}=U+W$ is explicitly written
as 
\begin{equation}
H_{C}=4E_{C}\left(S_{z}-n_{g}\right)^{2}+\mu n_{g}^{2}.\label{eq:HC}
\end{equation}
We remark that the last term $\mu n_{g}^{2}$ $=4E_{C}(C_{J}/C_{g}-1)n_{g}^{2}$
was neglected in some current references \cite{Nori2003,Wihelm2008},
since it is a constant for the classical voltage and not related with
the charges in the CPB. However, in the case of the gate voltage $V_{g}$
contains a quantized component, this term provides a nonzero quadratic
voltage term, which is evidently crucial in determining whether the
superradiant QPT exists.

\textit{Superradiant QPT in the Dicke model based on circuit QED.}\textemdash{}Now
we further consider the circuit QED system, as shown in Fig.$\ref{fig:geometry}$(b),
with a TLR coupled to $N$ small junctions, which are modeled as the
artificial atoms of two energy levels. The total Hamiltonian $H_{cir}=\sum_{j=1}^{N}H_{j}+\hbar\omega_{r}a^{\dagger}a$
is defined by 
\begin{eqnarray}
H_{j} & = & 4E_{C}\left(S_{z,j}-n_{g}\right)^{2}-E_{J}\cos\theta_{j}+\mu n_{g}^{2},\label{eq:Hto}
\end{eqnarray}
which correctly includes the tunneling part Eq.$\left(\ref{eq:HT}\right)$
and the charging part Eq.$\left(\ref{eq:HC}\right)$. Obviously, $H_{cir}$
is very similar to the cavity QED system for the atoms interacting
with cavity modes through minimum coupling. Generally, it is very
hard in experiments to realize the strong atom-field coupling in the
conventional cavity QED systems, but the strong coupling regime is
feasible in the current experiments of the superconducting circuit
QED. Therefore, the circuit QED system is more ideal to investigate
the superradiant QPT.

At the degenerate point the Josephson junction behaves as a two-level
system, and the total Hamiltonian reads 
\begin{eqnarray}
H_{cir} & = & \sum_{j=1}^{N}\left[E_{C}-\frac{E_{J}}{2}\sigma_{j}^{x}+\frac{2}{e}E_{C}C_{g}V_{q}\left(a+a^{\dagger}\right)\sigma_{j}^{z}\right]\nonumber \\
 &  & +\hbar\omega_{r}a^{\dagger}a+\hbar D\left(a^{\dagger}+a\right)^{2}+\hbar F\left(a^{\dagger}+a\right),\label{eq:Hcir}
\end{eqnarray}
where the correct two-photon term is included with $\hbar F=\mu NC_{g}V_{q}/2e$
and $\hbar D=NE_{C}C_{J}C_{g}V_{q}^{2}/e^{2}$. $\sigma_{j}^{x}=\left|1\right\rangle _{j}\left\langle 0\right|+\left|0\right\rangle _{j}\left\langle 1\right|$
and $\sigma_{j}^{z}=\left|0\right\rangle _{j}\left\langle 0\right|-\left|1\right\rangle _{j}\left\langle 1\right|$
are defined in terms of the two lowest eigenstates $\left|0\right\rangle _{j}$
and $\left|1\right\rangle _{j}$ of $S_{z,j}$. Here, we neglect a
constant $N\left(C_{J}/C_{g}-1\right)/4$ as it is a pure $c$ number
. 

To study the superradiant phenomenon in this circuit QED system, it
is necessary to explore the circumstance that the atom number $N$
is large and the total excitation number is low. In this case , the
collective excitation operator 
\begin{equation}
b^{\dagger}=\frac{1}{\sqrt{N}}\sum_{j}\left|e\right\rangle _{j}\left\langle g\right|\label{eq:b}
\end{equation}
behaves as bosonic operator in the atomic quasi-spin wave \cite{CPS2003},
which is defined by the eigenstates $\left|e\right\rangle _{j}=i\left(\left|0\right\rangle _{j}-\left|1\right\rangle _{j}\right)/\sqrt{2}$
and $\left|g\right\rangle _{j}=\left(\left|0\right\rangle _{j}+\left|1\right\rangle _{j}\right)/\sqrt{2}$
of the $j$-th CPB. Then, $H_{cir}$ is rewritten as 
\begin{eqnarray}
H_{cir} & = & \hbar\omega_{r}a^{\dagger}a+\hbar\omega_{J}b^{\dagger}b-i\hbar\Omega\left(a^{\dagger}+a\right)\left(b^{\dagger}-b\right)\nonumber \\
 &  & +\hbar D\left(a^{\dagger}+a\right)^{2}+\hbar F\left(a^{\dagger}+a\right).\label{eq:Hcir1}
\end{eqnarray}
Apparently, our circuit QED system is reduced into an equivalent system
of two coupled harmonic oscillators (CHO) with frequencies $\omega_{J}=E_{J}/\hbar$
and $\omega_{r}$, and the coupling strength is 
\begin{eqnarray}
\Omega & = & \frac{2\sqrt{N}E_{C}C_{g}V_{q}}{\hbar e}.\label{eq:Omega}
\end{eqnarray}

Generally, we consider a CHO system with canonical coordinates $x_{1}$
and $x_{2}$, eigenfrequencies $\omega_{1}$ and $\omega_{2}$, and
masses $m_{1}$ and $m_{2}$, respectively. If the coupling term were
inappropriately chosen as $-gx_{1}x_{2}$, the eigenvalues of the
coupled system would be imaginary when the coupling strength $g$
is strong enough, specifically, when $g^{2}>m_{1}m_{2}\omega_{1}^{2}\omega_{2}^{2}$.
Somebody depicts this phenomenon as a kind of QPT, but the natural
coupling in the conventional coupled CHO should be $g\left(x_{1}-x_{2}\right)^{2}/2=gx_{1}^{2}/2+gx_{2}^{2}/2-gx_{1}x_{2}$,
so the two quadratic coordinate terms renormalize the eigenfrequencies
as $\tilde{\omega}_{i}=\sqrt{\omega_{i}^{2}+g/m_{i}}$, $i=1,2$.
Therefore, there would not be QPT, since $g^{2}<m_{1}m_{2}\tilde{\omega}_{1}^{2}\tilde{\omega}_{2}^{2}$
is always valid \cite{CPSun2007}. This is the very reason that the
correct non-linear term is particularly important in the discussion
of QPT.

However, in our present circuit QED system, there is no such intrinsic
relation between the renormalized eigenfrequency and the effective
coupling, thus it is possible to observe such kind of QPT phenomenon
wherein. We would like to point out our model Hamiltonian contains
an additional linear term $\hbar F\left(a^{\dagger}+a\right)$, which
is introduced accompanying the correct two-photon term. It can be
eliminated by displaced transformations $\alpha=a+\eta$ and $\beta=b+\xi$
($\eta$ and $\xi$ are $c$ numbers). Then we diagonalize $H_{cir}$
in a conventional way \cite{Ripka1985} and obtain two eigenfrequencies
as 
\begin{equation}
\omega_{\pm}^{2}=\frac{1}{2}\left[\Omega_{+}^{2}\pm\sqrt{\Omega_{-}^{4}+16\Omega^{2}\omega_{r}\omega_{J}}\right],\label{eq:omega+-}
\end{equation}
where $\Omega_{\pm}=\sqrt{\omega_{r}^{2}+4D\omega_{r}\pm\omega_{J}^{2}}$.
It is obvious that $\omega_{+}^{2}$ is always positive, while the
$\omega_{-}^{2}$ can be negative when 
\begin{eqnarray}
1-\kappa\gamma & > & \frac{\omega_{r}\omega_{J}}{4\Omega^{2}}.\label{eq:condi}
\end{eqnarray}
Here, we define two dimensionless parameters, the ratio of Josephson
energy to charging energy $\kappa\equiv E_{J}/4E_{C}$ and the ratio
of Josephson capacitance to gate capacitance $\gamma\equiv C_{J}/C_{g}$.
It was proven in Ref.\cite{EmaryPRL} that $\omega_{-}=0$ is the
critical point of the superradiant QPT and the superradiant phase
lies in the region that eigenfrequency $\omega_{-}$ of the system
is imaginary, hence Eq.$\left(\ref{eq:condi}\right)$ is actually
the condition for the appearance of superradiant phase. As the coupling
$\Omega$ increases with $\sqrt{N}$, the right hand side of Eq.$\left(\ref{eq:condi}\right)$
approaches to positive infinitesimal in the limit $N\rightarrow\infty$.
Thus the occurrence of the superradiant QPT depends on the condition
$\kappa\gamma<1$. The corresponding critical point of $\Omega$ is
\begin{equation}
\Omega_{0}=\frac{1}{2}\sqrt{\frac{\omega_{r}\omega_{J}}{1-\kappa\gamma}}.\label{eq:Omega0}
\end{equation}

According to the above arguments, the superradiant QPT indeed can
occur in principle, but we need to examine this conclusion for the
realistic systems. In order to achieve a good charge qubit with small
fluctuation of the Cooper pair number and low classical noise, it
is usually chosen $\kappa\ll1$ and $\gamma\gg1$ in experiments.
These two key factors, $\kappa$ and $\gamma$, compete in determining
whether the superradiant QPT can take place. If the Josephson capacitance
is large enough, such as $\gamma\approx10$ \cite{Shnirman1999},
and the ratio $\kappa$ is set around $0.4$ \cite{Schoelkopf2004},
the condition Eq.$\left(\ref{eq:condi}\right)$ is violated thus the
superradiant QPT can not happen. In contrast, we can also choose $\kappa\ll1$
and $\gamma\approx1$ as in the experiment \cite{Delsing2005}, which
clearly allows the superradiant QPT.

\textit{Remarks and conclusion.}\textemdash{}Pedantically, we need
to understand why the Dicke-type superradiant QPT is allowed in the
circuit QED system, while it is forbidden in the cavity QED system,
since these two systems possesses very similar Hamiltonians with the
correspondences between canonical variables as listed in the table
\ref{tab:corrs}. This analogy apparently implies the superradiant
QPT can happen nether in the circuit QED system nor the conventional
cavity QED system. However, this argument obviously contradicts with
the conclusion made above, as well with the analysis by Nataf \textit{et
al} \cite{Nataf2011}.

\medskip{}

\begin{table}
\begin{tabular}{|c|c|}
\hline 
\multirow{1}{*}{natural atom} & artificial atom\tabularnewline
\hline 
$H_{na}=p^{2}/2m+U\left(x\right)$ & $H_{ar}=4E_{C}\left(S_{z}-1/2\right)^{2}-E_{J}\cos\theta$ \tabularnewline
\hline 
$x$ & $\theta$\tabularnewline
\hline 
$p$ & $S_{z}-1/2$\tabularnewline
\hline 
$\left[x,p\right]=i\hbar$  & $\left[\theta,S_{z}-1/2\right]=i$\tabularnewline
\hline 
\end{tabular}\caption{\label{tab:corrs}The correspondence between the natural atom and
the Josephson-type artificial atom.}
\end{table}

To solve this puzzle, we would like to consider whether or not there
exists the correspondence between the basis vectors used for defining
the collective operators in two systems. In cavity QED system, we
use the two lowest eigenstates $\left|e\right\rangle $ and $\left|g\right\rangle $
of the total Hamiltonian $H_{na}$ of a natural atom to define a qubit
subspace. Evidently, they are not the eigenstates of the momentum
operator $p$ due to the existence of trapping potential. In the artificial
atoms, however, though the electron pair number operator $S_{z}$
corresponds to momentum $p$, the two discrete eigenstates $\left|0\right\rangle $
and $\left|1\right\rangle $ of $S_{z}$ does not correspond to $\left|e\right\rangle $
and $\left|g\right\rangle $ respectively. It follows this observation
that the collective operators of the natural atom ensembles and artificial
atoms are of different types, and describe different types of quasi-excitations.
Thus it is not surprising that the circuit QED system exhibits superradiant
QPT while the cavity QED system does not.

In summary, we have theoretically explored the superradiant QPT in
the circuit QED system, where $N$ Cooper pair boxes behaves as artificial
atoms coupled to a single resonator mode. With the microscopic Hamiltonian
based on the pseudo-spin representation of the BCS theory and the
minimum coupling principle, we deduce  the correct quadratic term
$V_{g}^{2}$ of the gate voltage from a fully quantum perspective.
Then we showed that the circuit QED system is capable for the superradiant
QPT, and the critical point is determined by more parameters, the
ratios $\kappa$ and $\gamma$. The QPT is more feasibly to be realized
when these two ratios are small. We also explained the cavity and
circuit QED systems show different collective behaviors is due to
the superradiant phenomenons in these two systems are based on different
types of quasi-excitons.

\medskip{}

\begin{acknowledgments}
This work is supported by National Natural Science Foundation of China
under Grants No.11121403, No. 10935010 and No. 11074261.\end{acknowledgments}

\end{document}